%% file: MSR-2022.tex
\documentclass[sigconf]{acmart}

\usepackage{xspace}
\usepackage{color}
\usepackage{lineno}
\newif\ifdraft
\drafttrue

\input{macros}
\setcopyright{none}
\renewcommand\footnotetextcopyrightpermission[1]{}

\AtBeginDocument{%
  \providecommand\BibTeX{{%
    \normalfont B\kern-0.5em{\scshape i\kern-0.25em b}\kern-0.8em\TeX}}}

\nolinenumbers

\begin{document}

\title{Can instability variations warn developers when open-source projects boost?}
\titlenote{This study was accepted at the MSR 2022 Registered Reports Track.}

\author{Alejandro Valdezate}
\affiliation{%
  \institution{Universidad Rey Juan Carlos}
  \city{Madrid}
  \country{Spain}
  }
\email{alejandro@valdezate.net}

\author{Rafael Capilla}
\affiliation{%
  \institution{Universidad Rey Juan Carlos}
  \city{Madrid}
  \country{Spain}
}
\email{rafael.capilla@urjc.es}

\author{Gregorio Robles}
\affiliation{%
  \institution{Universidad Rey Juan Carlos}
  \city{Madrid}
  \country{Spain}
}
\email{gregorio.robles@urjc.es}

\author{Victor Salamanca}
\affiliation{%
  \institution{Universidad Rey Juan Carlos}
  \city{Madrid}
  \country{Spain}
  }
\email{victorsalamanca@gmail.com}

\renewcommand{\shortauthors}{Capilla et al.}

\begin{abstract}
Although architecture instability has been studied and measured using a variety of metrics, a deeper analysis of which project parts are less stable and how such instability varies over time is still needed.
While having more information on architecture instability is, in general, useful for any software development project, it is especially important in Open Source Software (OSS) projects where the supervision of the development process is more difficult to achieve. 
In particular, we are interested when OSS projects grow from a small controlled environment (i.e., the \emph{cathedral} phase) to a community-driven project (i.e., the \emph{bazaar} phase).
In such a transition, the project often explodes in terms of software size and number of contributing developers.
Hence, the complexity of the newly added features, and the frequency of the commits and files modified may cause significant variations of the instability of the structure of the classes and packages.
Consequently, in this registered report we suggest ways to analyze the instability in OSS projects, especially during that sensitive phase where they become community-driven.
We intend to suggest ways to predict the evolution of the instability in several OSS projects.
Our preliminary results show that it seems possible to provide meaningful estimations that can be useful for OSS teams before a project grows in excess. 
\end{abstract}




\maketitle

\section{Introduction}
\label{intro}
Software projects evolve over time to cope with changing requirements and maintenance operations. This evolution may cause an architectural mismatch between design and code, leading to architectural drift and erosion when the design diverges from the implementation or when a sub-optimal code violates architectural principles~\cite{Le2016}. Typically, software architectural mismatches are a consequence of design decisions~\cite{Capilla2017} that can impact negatively on the descriptive architecture (e.g., architectural drift) and on the prescriptive architecture (e.g., architectural erosion).

Today, Open Source Software (OSS) projects are no exception to the instability of the architecture. Given its nature, the risk of architecture instability is often higher as the design is seldom explicit~\cite{brown2011architecture} and the development team is heterogeneous and prone to a high developer turnover~\cite{lin2017developer}.
Even in industrial OSS projects, such as OpenStack, with a highly professionalized development team, having an overall picture is difficult to obtain, as developers come from many different companies, each with their own interests~\cite{teixeira2016cooperation,zhang2019companies}.
Among the OSS lifecycle, we have identified a phase where architecture instability is a major risk.
This happens when projects grow from a small size with a few developers to a community-driven project with hundreds of contributors~\cite{capiluppi2007cathedral}, where the activities follow a self-organized (stigmergic) pattern~\cite{robles2005self}.
During that transition phase, the automatisms among developers that were possible during the early stages of the project are not possible. And anyhow, becoming a community-driven project is a sign that the project attracts much interest, and that external effort, if conveniently integrated into the project, can set the project in another level~\cite{zhou2017scalability,tan2020scaling}.
It is in such scenarios where having metrics (and tools) on project instability that point out to risky parts in the source code would be very valuable.
Also, developers could be aware of parts of the project that need a further look and possibly action.
Additionally, other stakeholders, like external companies willing to invest in a project, would have information on the risk of having high architectural instability.

In this registered report, we propose to analyze several OSS projects, to identify those releases before becoming a community-driven project (i.e., to enter the \textit{bazaar} phase), and to evaluate them for architectural instability. Our aim is to find out which parts exhibit higher instability values hampering its evolution. We also plan to evaluate our findings against the feedback from project developers.

The structure of this registered report is as follows. Section 2 describes our motivation for this research work. In Section 3 we discuss some related work on instability metrics while Section 4 outlines the research questions. In Section 5 we detail the execution plan of the research and Section 6 reports the tools and datasets we plan to use. Finally, in Section 7 we discuss some initial results and in Section 8, potential implications for practitioners.


\section{Motivation}
Successful OSS projects (e.g., FreeBSD or Apache) often depend on large communities of developers sustaining the project releases~\cite{mockus2002two,dinh2005freebsd}. According to~\cite{capiluppi2007cathedral}, OSS projects start from a \textit{cathedral} phase where a small number of developers collaborate to achieve the main goal of the project. 
In the \textit{cathedral} phase, releases produce small-size software, offering the first evolution history of the project.
In this phase, although the architecture may change frequently, because of the small size of the development team and of the software, we can consider that the instability of the first releases does not vary in excess.
If the project achieves to attract the interest of other developers and users, and a significant number of developers engage into the project to add new functionality, the project ends up in the \textit{bazaar} phase.
The software architecture in this phase typically tends to stabilize as the project matures, as it has been reported for long-lived OSS projects~\cite{gonzalez2014studying}.
We hypothesize that it is in the transition between the \textit{cathedral} and the \textit{bazaar} phases where the instability of the releases may be the result of an increasing number of changes and contributors.

\begin{figure}[ht!]
\begin{center}
\includegraphics[keepaspectratio=true,width=0.42\textwidth]{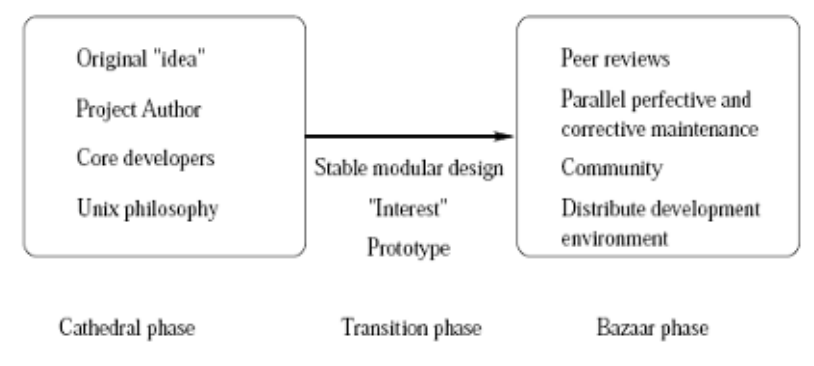}
        \caption{Transition from the \emph{cathedral} phase to a \emph{bazaar} phase. Taken from~\cite{capiluppi2007cathedral}.}
        \label{fig:_catB}
\end{center}
\end{figure}


Figure~\ref{fig:_catB} illustrates three basic phases, which~\cite{capiluppi2007cathedral} argues a successful OSS project undergoes.
The initial phase of an OSS project operates in the context of a community of volunteers.
All the characteristics of the cathedral style development (like requirements gathering, design, implementation and testing) are present, and they are carried out in the typical style of building a cathedral. In this case, the work is done by an individual or a small team working in isolation from the community~\cite{bergquist2001power}.
This development process shows tight control and planning from the central project author, and is referred to as `closed prototyping' by Johnson~\cite{johnson2001descriptive}.

In order to become a high quality and useful product, \cite{senyard2004have} argued that an OSS project has to make a transition from the cathedral phase to the bazaar phase (as depicted by the arrow in Figure~\ref{fig:_catB}). 
In this phase, users and developers continuously join the project writing code, submitting patches and correcting bugs.
This transition is associated with many complications: it is argued that the majority of OSS projects never leave the cathedral phase and therefore do not access the vast amount of resources of manpower and skills the OSS community offers~\cite{munaiah2017curating}. Also, knowing the behavior of the instability over time is of major interest for companies contribution with development effort, as projects significantly start growing is an optimal time and companies would like acquire an important technological advantage for the future~\cite{gonzalez2013understanding}. 

\begin{figure*}[ht!]
\begin{center}
\includegraphics[keepaspectratio=true,width=0.9\textwidth]{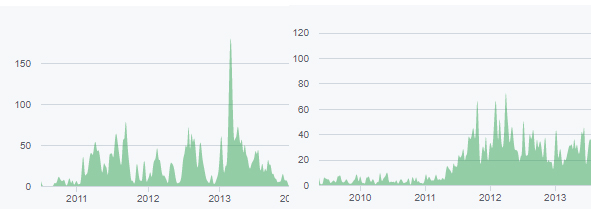}
        \caption{Evolution of number of committers (per week) for Catroid (left) and Hadoop (right). Source: GitHub contributors graphs.}
        \label{fig:Catroid}
\end{center}
\end{figure*}

The shift between phases is often risky because when projects increase significantly the number of developers and some control is lost when releases are more and more distributed among the different OSS teams. For instance, Figure~\ref{fig:Catroid} shows the evolution in number of weekly committers for the Catroid and Hadoop projects, as offered by GitHub. For instance, for Catroid we can observe this shift between the cathedral and bazaar phases in 2011-2012 (although there is a second peak in 2013-2014). For Hadoop, the transition phase can be observed during 2011. Before it, Catroid and Hadoop count with a reduced group of participants (around 5); after it, the number of contributors is always above 20 with peaks well above 50.



\section{Instability metrics}

\subsection{Instability metrics in packages}
Software architecture packages are high-level entities commonly used to describe subsystems or to group related functionality. In many complex systems, some packages depend on others. Bouwers analyzes the modularity and module dependencies to infer maintainability problems in software architectures~ \cite{Bouwers2009}. In line with Bouwers, Koziolek~\cite{Koziolek2011} suggests more than 40 metrics that investigate the sustainability of an architecture \cite{Lago2015, Penzenstadler2013, Venters2014} measuring the dependencies between modules among other metrics. These dependencies can be used to infer the instability of architectural elements. For instance, the Linux system often requires additional packages when installing and configuring new functionality. As a consequence, a set of dependencies is established between packages and the modifications in one of these packages affect other related packages. Some works (e.g.,~\cite{Alves2011}) perform a comparison of code querying languages and tools based on an implementation of the instability metric defined in~\cite{MAR94} according to the number afferent coupling (i.e., classes outside a package that depend on classes inside the package) and efferent coupling (i.e., classes inside a package that depend upon classes outside the package).

Another work~\cite{Alenezi2015} suggests ways to estimate package stability metrics to measure the changes affecting the stability of the architecture and changes that happen during system evolution. The authors estimate the instability of two consecutive releases from these changes and provide an aggregate measure of the system instability. In addition, the authors in~\cite{BAIG2019} suggest a package stability metric (PSM) based on the changes between package contents and the relationships inside the package. The proposed metric estimates the maintenance effort and computes package stability based on three dimensions: content, internal package connection, and external package connections~\cite{Alshayeb2010}. 

More recently, in~\cite{Fontana2017} the authors mention that a package is less stable if it depends on an unstable related package. They suggest a metric called ``Degree of Unstable Dependency" as the ratio between the number of dependencies that makes a package unstable (i.e., BadDependency) and the total number of dependencies. Finally, in~\cite{BAIG2019}, the authors describe a new package stability metric based on the changes between package contents and intra- and inter-package connections that validate empirically in five OSS programs. The authors found a negative correlation between the proposed metric and the maintenance effort and a positive correlation between the package stability metrics based on changes in lines of code and class names.

\subsection{Instability metrics in classes}
One of the seminal works suggesting the idea of stability in object-oriented (OO) design was~\cite{MAR94}, who stated a way to measure the instability between OO classes. Also, the authors in~\cite{LI2000} suggest three different instability metrics named System Design Instability (SDI), Class Implementation Instability (CII), and System Implementation Instability (SII), which are used to estimate the evolution of OO systems via the analysis of the instability of object-oriented designs and classes that changed. Regarding the SDI metric, the authors conclude that the instability of the project examined is higher in the early stages of the development phase. In addition, in~\cite{Rapu2004}, the authors consider a class is stable with respect to a measurement of a previous version if there is no change in that measurement. They provide a formula to estimate the stability of classes applied to a class history, which is computed as the fraction of the number of versions in which a class changes over the total number of versions. 

Nowadays, one of the principal indicators of architecture erosion is the instability of the architecture or a system. Instability is an indicator that analyzes the frequency of changes of a system over time. According to ~\cite{MAR94}, instability uses the number of dependencies between elements to discover the effect of changes. One estimator of instability is the ripple effect of changes~\cite{Black01}, as there is a strong correlation between the ripple effect and the logical stability of software modules based on the cascading impact of changes. 


The majority of the studies discussed above provide coarse-grained instability estimations and do not study in depth the evolution of the instability across parts of OSS projects. Hence, the aim of this registered report is twofold: (i) to study the variations of instability values in different parts of the code of several OSS projects when we add or remove classes, and (ii) to investigate how instability could be predicted in order to warn developers about significant variations.

\section{Research questions}
We have been inspired for this registered report by a previous work~\cite{Carrillo2018} that describes an instability metric to estimate the ripple effect of design decisions.
We will use the formula described in~\cite{MAR94} to compute the instability values of OSS projects.
To advance the state of the art, we will investigate which parts of OSS projects exhibit more instability and how we can assess developers about instability variations.
With this aim, we will conduct an exploratory case study~\cite{Runeson2009,Yin2014} in several OSS projects to uncover the estimation and evolution of instability measures. 
We will address the following research questions:

\begin{itemize}
\item {\bf RQ1. Can we predict instability variations along the evolution of OSS projects?}\\
\textbf{Rationale:} In this research question, we attempt to provide trends of the evolution of the architectural instability and use these results to analyze which projects are more stable. We also attempt to predict future instability changes, based on the ratio of modified files. In this research question we plan to analyze evolution trends of the instability values, and identify hot-spots based on the number of modified files that can suggest future variations of the instability values. Additionally, we expect to identify the most promising periods where more developers are needed and use this data as input for RQ2.
\item {\bf RQ2. How do new functionality, bugs and refactorings affect the instability in OSS projects when they shift from the \emph{cathedral} to the \emph{bazaar} phase?}  \\
\textbf{Rationale:} Changes to the project may affect the architectural instability of the project, especially if most of these changes are based on adding and removing classes and relationships between classes. Our hypothesis is that during the transition from the \emph{cathedral} to the \emph{bazaar} this occurs frequently. Therefore, in this research question we will investigate how such changes between project releases modify the instability values during the transition period. The expected outcomes are i) to understand how instability varies inside a release when packages are added and removed, and ii) to identify which packages (from a subset of releases) exhibit more instability, in both chases during the transition period.
  \end{itemize}

\section{Execution Plan}
According to the ACM guidelines~\cite{Ralph2021}, we will follow an exploratory case study for the experiment design. We will select releases and packages of several OSS projects described in the dataset section. To apply the instability metrics, we will compute the instability of each project at the class level and their dependencies. Hence, we will adopt following protocol:
\begin{enumerate}
    \item  We will select a set of OSS projects where we can identify a transition from the \emph{cathedral} to the \emph{bazaar} phase. Thus, we have to define three aspects: i) identify the cathedral phase, ii) identify the bazaar phase, and iii) specify the time interval for the change.

        \begin{itemize} 
         \item As for i) and ii), we have looked at the scientific literature for any type of definition in this regard, but have not found anything. Our position is that this can be done merely based on the number of committers in a given time period. As this has not been previously researched, we propose two tentative numbers that we find reasonable at this point: we expect for the cathedral phase less than 10 committers in a month, while for the bazaar phase it should be more than 50 committers in a month.

         \item As for iii), we think that a reasonable time span is in the range of 12 to 24 months. We will therefore start looking for projects where conditions i) and ii) apply in 12 months, using a sliding window algorithm. If we find that this timespan in too short, we will try with timespans of 18 months and 24 months, respectively.

        \item To offer some visual evidence of our decision, we have taken two projects as examples.
Figure~\ref{fig:Catroid} shows the evolution of committers (on a weekly basis) to the Catroid and Hadoop projects, respectively, as taken from GitHub. We can observe that in both cases around 2011 there is a transition phase between the cathedral and the bazaar. We also can observe how in the case of Hadoop the high activity has been maintained since then, while for Catroid there is more variance.
      \end{itemize}
      
    \item We will attempt to provide a way to predict variations in the instability values and estimate where more developers have more activity in the project, according to the following sub-steps:
     \subitem (a) First, we will compute the ratio of modified files for a given time frame, independently of the number of the commits. 
     \subitem (b) Second, we will calculate the ratio of modified files alongside the instability values for different project releases, such as the ones shown in Figure~\ref{fig:Insta-files} for four sample projects.
     \subitem (c) Third, we will manually analyze the trends where the ratios of modified files are bigger to delimit the periods where more developers participate and we will identify future variations of instability ratios in subsequent releases.
    
    \item We will analyze the instability for different releases of several OSS projects to investigate the differences of instability values when new functionality is added. For this aim we will perform the following sub-steps:
        \subitem (a) According to the cathedral and bazaar phases, we will select those periods where we observe a significant activity of developers. 
        
        \subitem (b) We use the ARCADE tool we will obtain the dependencies between the classes of each of the releases selected. For instance, according to Figure~\ref{fig:Catroid} we can select two different periods (i.e. 2011-2012 and 2013-2014), where the number of developers boosts.
        \subitem (c) We will transform the data obtained from ARCADE to a format that can be read by algorithm we developed to compute the instability values of the releases.  
       \item We will apply an statistical analysis of the results using the Pearson test to find possible correlation between the instability values and the independent variables. In case the Pearson test doesn't provide significant results we will use other statistical tests.
  \end{enumerate}

\section{Tools and datasets}

\subsection{Tools}
We will use the ARCADE\footnote{\url{https://bitbucket.org/joshuaga/arcade}} tool~\cite{Laser2020}, a workbench for software architecture recovery and decay analysis that can retrieve different architectural models and visualize architecture decay using different metrics to detect smells. Ideally, ARCADE can be used to study the differences and evolution of software architectures and provide data about architectural changes. ARCADE can discover the dependencies between classes, as we will use these dependencies in the instability formula. 

For our study, we will need to download the specific version of the project and find all the libraries involved in order to generate the jar for compilation. Then, we will analyze the compiled version using ARCADE. Since ARCADE runs on top of an Oracle JVM, the analysis is limited by the memory that this virtual machine (VM) allocates for running JAVA programs. Therefore, a limit to the selection of the target projects is its size, as larger projects may result in buffer overflows of the VM and make the analysis not viable.

\subsection{Datasets}

%

To investigate the instability in real OSS projects, we will select OSS projects that comply with following criteria:

\begin{enumerate}
  \item Have a repository in GitHub and are not forks
  \item Are written in Java
  \item A transition phase between cathedral and bazaar phases can be identified
  \item Are compilable
  \item Can be analyzed using the ARCADE tool
  \item We plan to interview developers from selected open-source projects to double-check our results
\end{enumerate}

Our goal is to have at least six projects that fit these criteria.
The number of projects is limited by the fact that we want to obtain feedback from their developer community by means of interviewing them.
The rationale for this is that we want to gain deep knowledge on a smaller set of projects that can serve as case studies, instead of obtaining superficial knowledge from a statistical study of many OSS projects.

According to Bird et al.~\cite{bird2012assessing} we refer to the importance of detailed studies of few projects.
The authors recall that it is a misconception that this type of studies, on a small number of projects, do not provide to the academic community any value, nor contribute to scientific development.
In the final version of our work, we will discuss why our findings can (or not) be representative of many other OSS projects.

A replication package will contain the list of projects, and annotations with our manual inspections, and transcription of the interviews (probably in an anonymized way).

%
%
%
%
%
%
%

\section{Preliminary analysis}

\begin{figure*}[ht!]
\begin{center}
\includegraphics[keepaspectratio=true,width=0.93\textwidth]{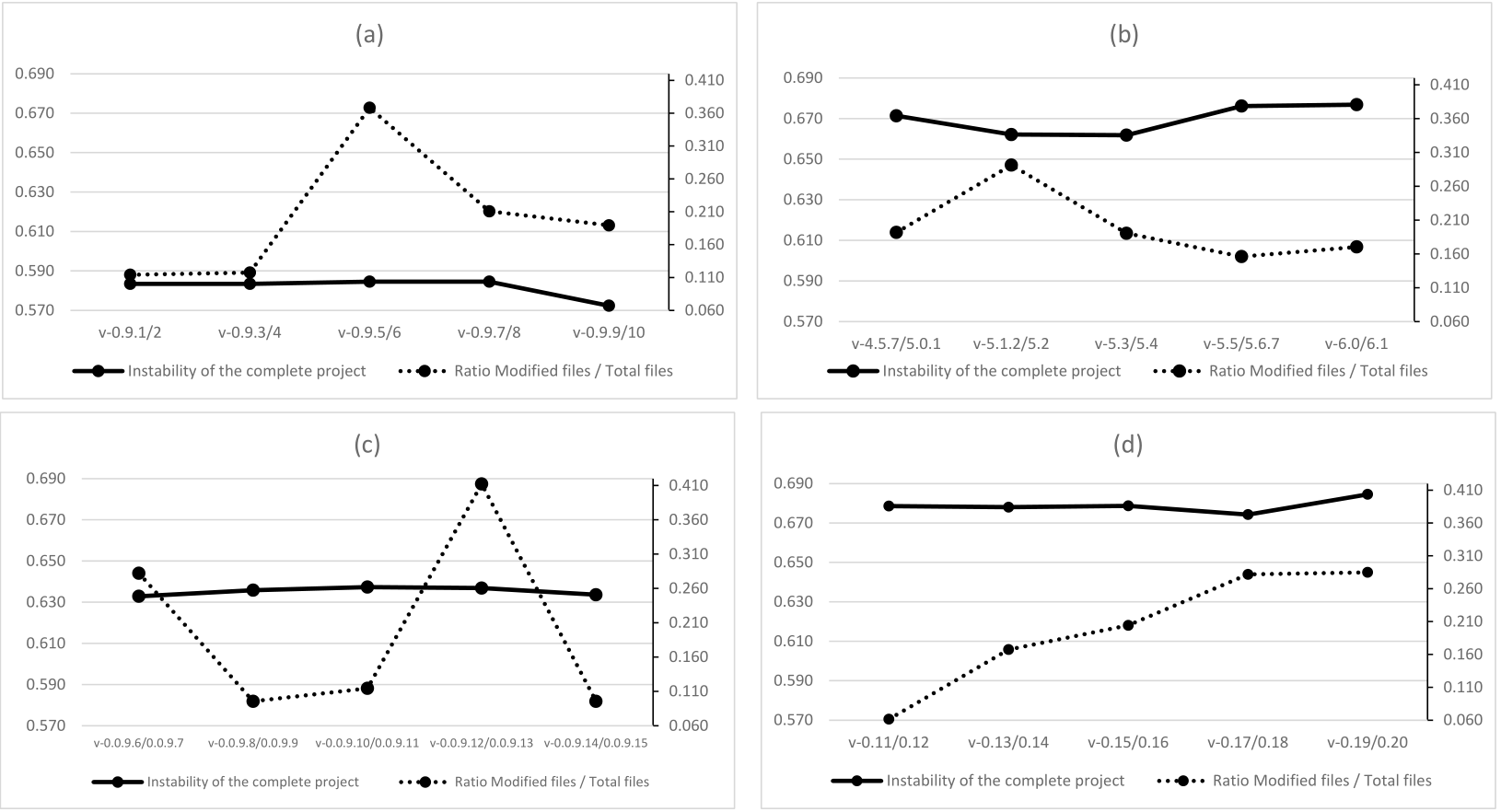}
        \caption{Comparison between Instability and ratio of the modified files for (a) Catroid, (b) SonarQube, (c) dex2jar and (d) Hadoop.}
        \label{fig:Insta-files}
\end{center}
\end{figure*}

To show the feasibility of our approach, we selected four OSS projects (Catroid\footnote{\url{https://github.com/Catrobat/Catroid}}, SonarQube\footnote{\url{https://github.com/SonarSource/sonarqube}}, dex2jar\footnote{\url{https://github.com/pxb1988/dex2jar}} and Hadoop\footnote{\url{https://github.com/apache/hadoop}}), and we performed a preliminary analysis on their instabilities for a timeframe in order to show how they evolve. According to Figure~\ref{fig:Insta-files}, we can observe the instability trends for the four projects analyzed and the ratios of the modified files. 

Based on this information, we need to connect these ratios with the periods belonging to the cathedral and bazaar phases to estimate instability changes. These initial results can serve to identify more unstable packages as well as developers working on them, and provide fine-grained instability indicators.
For completeness, we will contact developers to gain further insight into the history of the project and to ascertain if our method and results are relevant to understand the transition phase.


Although it is hard to provide at this stage take-away lessons for practitioners, we believe our preliminary results can provide a deeper understanding of the instability evolution and changes along specific time frames. We use the selected shifts between the cathedral and bazaar phases to uncover what project branches exhibit higher instability values. These results can be used by open-source communities and other stakeholders (such as companies wanting to invest in the software project) to identify risks when they have a steady growth.

\section{Contributions and implications}

Our main contribution is threefold. First, researchers would better understand how the instability of OSS projects evolve, and could be able to evaluate which packages behave in a more unstable way. Additionally, we can suggest ways to identify instability hot-spots as indicators of future variations of the instability for different releases, and use these hot-spot as predictors of instability changes. Also, we attempt to investigate the effect in the instability of each release of classes added and removed, and use these to analyze if small variations of the instability values between releases can suggest a trend. 

Second, practitioners and developers may use these instability estimators to analyze their behavior in each package, and understand what parts of the software are becoming more unstable. We can use the insights as well to see what developers induce more instability, and try to find bad programming practices and their underpinning reasons.    

Third, understanding the evolution of instability in the transition period from the \emph{cathedral} to the \emph{bazaar} can help managers and business angels wanting to invest in an OSS project to evaluate the project and its risks. The analysis can provide valuable knowledge to adopt measures against this instability, as well.









\bibliographystyle{ACM-Reference-Format}
\bibliography{references}

\appendix





\end{document}

%% file: macros.tex
\ifdraft
  \newcommand{\grex}[1]{{\color{blue}\emph{Gregorio says: #1}}\xspace}
  \newcommand{\rafa}[1]{{\color{red}\emph{Rafael says: #1}}\xspace}
  \newcommand{\carlos}[1]{{\color{orange}\emph{Carlos says: #1}}\xspace}
  
  \newcommand{\fixme}[1]{{\textcolor{red}{[FIXME] #1}}\xspace}
  
\else
  \usepackage[disable]{todonotes}
  \newcommand{\grex}[1]{}
  \newcommand{\rafa}[1]{}
  \newcommand{\carlos}[1]{}
  \newcommand{\fixme}[1]{}

\fi

\usepackage[inline]{enumitem}

{\begin{center}\vspace{1mm}\noindent\begin{Sbox}\begin{minipage}{0.95\columnwidth}}%
{\end{minipage}\end{Sbox}\fbox{\TheSbox}\end{center}\vspace{1mm}}
